# Intelligent DoS and DDoS Detection: A Hybrid GRU-NTM Approach to Network Security

Caroline Panggabean
*Departement of CSE (AI)*
JAIN (Deemed – to – be University)
Bangalore, Karnataka
carolinepgabean@gmail.com

Chandrasekar Venkatachalam
*Departement of CSE (AI)*
JAIN (Deemed – to – be University)
Bangalore, Karnataka
chandrasekar.v@jainuniversity.ac.in

Priyanka Shah
*Departement of CSE (AI)*
JAIN (Deemed – to – be University)
Bangalore, Karnataka
priyankashah8324@gmail.com

Sincy John
*Departement of CSE (AIM)*
JAIN (Deemed – to – be University)
Bangalore, Karnataka
sincyjohn@jainuniversity.ac.in

Renuka Devi P
*Departement of CSE (AIML)*
JAIN (Deemed – to – be University)
Bangalore, Karnataka
renukadevi.p@jainuniversity.ac.in

Shanmugavalli Venkatachalam
*Department of CSE*
KSR College of Engineering
Namakkal, Tamil Nadu
drvshanmugavalli@gmail.com

*Abstract*— Detecting Denial of Service (DoS) and Distributed Denial of Service (DDoS) attacks remains a critical challenge in cybersecurity. This research introduces a hybrid deep learning model combining Gated Recurrent Units (GRUs) and a Neural Turing Machine (NTM) for enhanced intrusion detection. Trained on UNSW-NB15 and BoT-IoT datasets, the model employs GRU layers for sequential data processing and an NTM for long-term pattern recognition. The model achieves 99% accuracy in distinguishing between normal, DoS, and DDoS traffic. This approach offers promising advancements in real-time threat detection, potentially improving network security across various domains.

*Keywords— Intrusion Detection System, Gated Recurrent Unit, Neural Turing Machine. DoS, DDoS*

## I. Introduction

Detecting network intrusions is a constant struggle in cybersecurity, especially with the growing complexity and frequency of Denial of Service (DoS) and Distributed Denial of Service (DDoS) attacks. These threats pose serious risks to network systems and services. Traditional methods, which rely on predefined signatures to detect attacks, often struggle to keep pace with new and evolving attack patterns.

While machine learning approaches have shown promise, they often struggle with the temporal nature of network traffic and the need for long-term memory. Current models may excel at pattern recognition but fall short in understanding the context and evolution of attack strategies over time.

Our research addresses critical security concerns in modern networks, including volumetric DoS and DDoS attacks that overwhelm network resources, application-layer attacks mimicking legitimate traffic, slow and low attacks that gradually degrade network performance, and the challenge of distinguishing between flash crowds and DDoS attacks. We also consider the threat of zero-day attacks with previously unseen patterns. This study aims to develop a hybrid GRU-NTM model capable of accurately distinguishing between normal traffic, DoS, and DDoS attacks; evaluate the model's performance against advanced intrusion detection systems; and analyse the model's ability to adapt to evolving attack patterns.

Our proposed GRU-NTM hybrid model uniquely combines the sequential learning capabilities of GRUs with the flexible, long-term memory of Neural Turing Machines. This synergy allows for both immediate pattern recognition and the ability to reference and update information over extended periods, mirroring the complex, time-dependent nature of network attacks. Improving DoS and DDoS detection could significantly enhance network stability and reduce service downtime. For businesses, this translates to improved customer trust, reduced financial losses, and maintained productivity. On a broader scale, it contributes to a more secure and reliable digital infrastructure essential for our increasingly connected world.

Our research bridges the gap between deep learning architectures and network security, leveraging advances in neural network design to address critical cybersecurity challenges. This interdisciplinary approach combines insights from computer science, data analytics, and network engineering to create a more robust defence against modern cyber threats.

Developing such a system presents several challenges, including the need for extensive computational resources, the complexity of training a hybrid model, and the difficulty of creating representative datasets that capture the diversity of real-world network traffic and attack scenarios. Our research addresses these challenges through innovative model design and careful data preprocessing techniques.

## II. Related Work

Recent progress in the detection of Denial of Service (DoS) and Distributed Denial of Service (DDoS) attacks has primarily centered on the application of machine learning (ML) and deep learning (DL) methodologies. These approaches aim to improve accuracy, reduce false positives, and enhance real-time detection capabilities.

Several studies have explored hybrid models combining different ML algorithms. Coscia et al. [1] proposed Anomaly2Sign, an algorithm that automatically generates Suricata rules using a Decision Tree-based approach, achieved high classification metrics (99.7%-99.9%). This approach outperformed traditional classifiers like Logistic Regression and Support Vector Machines. Alfatemi et al. [2] combined Deep Residual Neural Networks with synthetic oversampling, demonstrating a remarkable accuracy of 99.98% on the CICIDS dataset, addressing the common issue of class imbalance in cybersecurity datasets.

Convolutional Neural Networks (CNNs) and Long Short-Term Memory (LSTM) networks demonstrated the potential to detect spatial and sequential patterns in network traffic. Issa and Albayrak [3] proposed a hybrid CNN-LSTM model that achieved 99.20% accuracy on the NSL-KDD dataset, surpassing previous works. Similarly, Al-zubidi et al. [4] introduced a CNN-LSTM-XGBoost model that achieved high accuracy across multiple datasets: 98.3% for CICIDS-001, 99.2% for CICIDS2017, and 99.3% for CIC-ID2018, demonstrating the model's robustness across diverse datasets.

Feature selection and dimensionality reduction have been crucial in improving model performance. Mebawondu et al. [5] used gain ratio for attribute ranking and selected the top 30 attributes for their Artificial Neural Network-based IDS, achieving 76.96% accuracy on the UNSW-NB15 dataset. This approach demonstrated the potential of lightweight IDS for real-time intrusion detection. Zeeshan et al. [6] introduced a Protocol Based Deep Intrusion Detection (PB-DID) effectively reduced the feature set while achieving a 96.3% accuracy, addressing issues of imbalance and overfitting in public datasets.

The integration of Software-Defined Networking (SDN) with IoT has created new opportunities for DDoS detection. Bhayo et al. [7] shown a machine learning-based architecture for DDoS attack detection in SDN-WISE IoT controllers, achieving accuracy rates of up to 98.1% using Decision Trees. Their framework integrated ML algorithms directly into the SDN-WISE controller for efficient packet classification. Ali et al. [8] compared various ML/DL approaches in SDN environments, finding that Support Vector Machines (SVMs) demonstrated the highest prediction accuracy (95.5%), while CNNs showed high training accuracy but lower prediction accuracy.

Cloud computing security has also benefited from ML-based DDoS detection approaches. Mohammed [9] evaluated various classifier models, including Random Forest, SVM, and Multi-Layer Perceptron (MLP), with MLP achieving a remarkable accuracy of 99.8% in detecting DDoS attacks in cloud environments. This study emphasized the importance of feature selection and normalization in enhancing model performance.

Smart home networks, being particularly vulnerable to DDoS attacks, have also been a focus of recent research. In Garba et al. [10], a framework for real-time DDoS attack detection and mitigation in SDN-connected smart homes was proposed. Their study found that the Decision Tree algorithm excelled in attack detection with an accuracy of 99.57%, demonstrating the potential of ML in protecting IoT devices.

Recent research has also explored the potential of quantum computing in enhancing ML performance for DDoS detection. Said [11] introduced a quantum support vector machine (QSVM) model that demonstrated better accuracy and computational resource efficiency compared to classical SVM models, opening up new possibilities for cybersecurity in the quantum computing era.

As the field advances, researchers are focusing on real-time detection capabilities and addressing the challenges of evolving attack patterns. Berei et al. [12] achieved a 99% success rate in detecting cyberattacks in real-time environments using ML models trained on a double-feature-reduced dataset. This study emphasized the importance of feature reduction in improving model efficiency and precision.

Ensemble methods have shown promise in improving detection accuracy. Das et al. [13] proposed an ensemble-based approach combining supervised and unsupervised learning frameworks, achieving up to 99.1% accuracy in detecting DDoS attacks across multiple datasets. This approach demonstrated the potential of combining different ML paradigms to enhance detection capabilities.

In the context of Voice over IP (VoIP) security, Lina et al. [14] reviewed deep learning techniques, specifically CNNs and RNNs, for detecting DDoS attacks. Their study found that these techniques achieved F1-scores above 96% in VoIP environments, highlighting the adaptability of deep learning methods to different network contexts.

Al-Eryani et al. [15] conducted a comparison of various ML algorithms for DDoS detection using the CICDoS2019 dataset. Their research found that ensemble methods, particularly Gradient Boosting (GB) and XGBoost, outperformed other algorithms, with GB achieving 99.99% accuracy and XGBoost 99.98% accuracy, along with low false alarm rates.

Despite significant advancements in DoS and DDoS attack detection using machine learning and deep learning techniques, several limitations persist. One major issue is the reliance on outdated or synthetic datasets, which may not accurately represent real-world attack scenarios [6][14]. Additionally, while many models demonstrate high accuracy in controlled environments, their performance in real-time, dynamic network conditions remains largely untested [10][12]. This raises concerns about their adaptability to new attack patterns, as most models are trained on known patterns and may struggle to detect novel or evolving threats [13][15].

Another limitation is the substantial computational resource requirements of some advanced models, particularly those employing deep learning approaches, which can hinder their practical implementation [2][4]. Furthermore, the interpretability of these high-performing models is often lacking, making it difficult to understand the reasoning behind their decisions [3][14]. Class imbalance in cybersecurity datasets also poses a challenge, potentially leading to biased models [2][6].

Moreover, many studies evaluate their models on a single dataset, raising questions about their generalizability [5][8]. Scalability issues are another concern, especially as network sizes and complexities grow in IoT and cloud environments [7][9]. These limitations underscore the need for ongoing research and development to enhance the effectiveness and effectiveness of ML and DL techniques in DoS and DDoS attack detection.

TABLE I.    EXISTING RESEARCH

| Authors | Technique(s) | Accuracy | Key Findings |
|---|---|---|---|
| [1] Coscia et al | Decision Trees, Anomaly2Sign | 99.7% - 99.9% | High accuracy, effective in handling large-scale datasets |
| [2] Alfatemi et al | Deep Residual Neural Networks with SMOTE | Up to 99.9% | Addressed class imbalance, high detection rates, low false alarms |

| Authors | Technique(s) | Accuracy | Key Findings |
|---|---|---|---|
| [3] Issa, A.S.A., & Albayrak, Z. | CNN-LSTM Hybrid | 99.20% | Outperformed individual CNN and LSTM models |
| [4] Al-zubidi, A.F., et al. | CNN-LSTM-XGBoost | 98.3% - 99.3% | High performance across multiple datasets |
| [5] Mebawondu, J.O., et al. | ANN-MLP with Gain Ratio | 76.96% | Lightweight IDS suitable for real-time detection |
| [6] Zeeshan, M., et al. | Deep Learning (PB-DID) | 96.3% | Reduced features, comprehensive coverage of benchmark datasets |
| [7] Bhayo, J., et al. | ML in SDN-WISE (DT) | 98.1% | Effective in SDN-IoT environments |
| [8] Ali, T.E., et al. | SVM | 99.8% | Best performance among various ML/DL approaches in SDN |
| [9] Mohammed, A. | MLP | 99.57% | Highest accuracy among evaluated models |
| [10] Garba, U.H., et al. | Decision Tree | Above 96% F1-score | Effective in smart home networks |
| [11] Said, D. | QSVM, Quantum Computing | Not Specified | Showed potential for enhanced DDoS detection using quantum computing |
| [12] Berei, E., et al. | ML Models, Double-Feature Reduction | 99% | High accuracy in real-time environments |
| [13] Das, S., et al. | Ensemble (Supervised + Unsupervised) | Up to 99.1% | Effective in detecting both known and novel attacks |
| [14] Lina, B., et al. | CNN, RNN | F1 > 96% | Promising for VoIP environments |
| [15] Al-Eryani, A.M., et al. | Gradient Boosting, XGBoost | 99.99% (GB), 99.98% (XGBoost) | high accuracy and low false alarms. |

## III. PROPOSED WORK

### A. Proposed Model

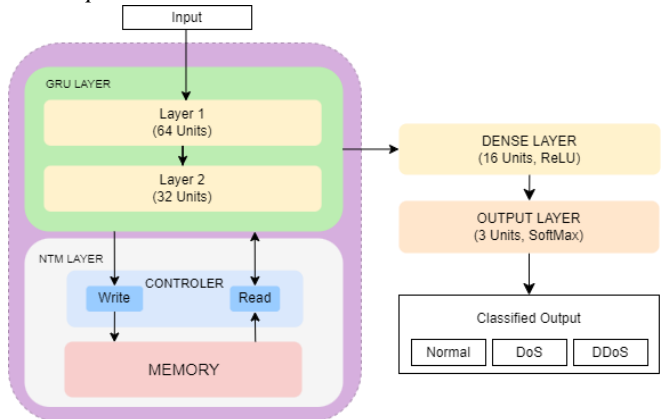

Fig. 1. Architecture Diagram

The proposed architecture combines a Gated Recurrent Unit (GRU) with a Neural Turing Machine (NTM) to create a powerful hybrid model for detecting Denial of Service (DoS) and Distributed Denial of Service (DDoS) attacks. The model's structure can be broken down into several key components:

*1) Input Layer:* The input layer receives the preprocessed network traffic data, which has been normalized and structured into appropriate time windows.

*2) Gated Recurrent Unit (GRU):* The GRU layers, chosen for their efficiency in processing sequential data, capture short-term dependencies in network traffic patterns. It consists of two layers:

- *GRU Layer 1:* 64 units. This layer processes the data sequentially, capturing short-term dependencies and patterns in the traffic data.
- *GRU Layer 2:* 32 units. The output from the first GRU layer, which includes processed sequences, is fed into the second GRU layer with 32 units. This layer further refines the temporal dependencies, enhancing the model's understanding of the sequence data.

*3) Neural Turing Machine (NTM):* The NTM component complements the GRU layers by maintaining long-term memory. It consists of a controller (implemented as GRU layers), an external memory matrix, and read/write heads. This structure allows the model to maintain context over long sequences of network traffic, crucial for detecting complex, evolving attack patterns.

- *Memory:* A key feature of the NTM is that it acts as an external storage and retrieval mechanism, enabling the model to capture long-term dependencies in network traffic patterns. It allows the model to store relevant information about past traffic behaviors and access it when needed, which is crucial for detecting evolving or persistent attack patterns. The memory interacts with the GRU layers through read and write operations, enabling the model to maintain context over extended periods while processing sequential data. This feature is particularly valuable for identifying sophisticated DoS and DDoS attacks that may develop gradually or exhibit complex temporal patterns, significantly improving the model's overall detection capabilities and adaptability to various attack strategies.
- *Read and Write Operations:* The read operation allows the model to retrieve information from memory. During this process, the controller, implemented as GRU layers, generates read vectors that act as queries to the memory. These queries are used to compute attention weights over the memory locations, and the weighted sum of memory contents, based on these attention weights, is returned as the read output. This mechanism enables the model to selectively focus on relevant past information when processing current network traffic, aiding in the detection of complex, time-dependent attack patterns. On the contrary, the write operation updates the memory with new information. The controller produces write vectors containing new information to be stored and erase vectors determining which parts of the existing memory should be cleared. This process allows the model to dynamically update its knowledge base, storing new patterns of network behavior and potentially overwriting outdated information. Such adaptive memory management is crucial for maintaining an up-to-date understanding of evolving DoS and DDoS attack strategies.

*4) Dense Layer:* After the GRU-NTM processing, the output is fed into a dense layer with 16 units using ReLU activation. This layer helps in further feature abstraction and non-linear combinations of the learned representations. This bridges the sequential processing of the GRU-NTM with the final classification, reducing dimensionality and allowing feature interactions. This layer enhances the model's ability to detect complex patterns in network traffic, enhancing its effectiveness in distinguishing between normal, DoS, and DDoS traffic. It provides a flexible point for capacity control and helps prevent overfitting, ultimately contributing to the model's generalization capability in detecting network attacks.

*5) Output Layer:* The final layer consists of 3 units with SoftMax activation, corresponding to the three possible classifications: Normal, DoS, and DDoS.

*6) Classified Output:* The model produces a probability distribution over the three possible classes, allowing for a nuanced interpretation of the network traffic classification.

This architecture leverages the strengths of both GRUs and NTMs:

- The GRU layers excel at processing sequential data and capturing short to medium-term dependencies in network traffic.
- The NTM's external memory allows the model to store and retrieve important information over long sequences, which is crucial for detecting sophisticated attack patterns that may evolve over time.
- The combination of these elements enables the model to maintain context over long periods while still being able to quickly adapt to new patterns in the input data

Detection is achieved by analyzing traffic patterns over time using GRU layers. It identifies distributed patterns across multiple sources through memory capabilities of Neural Turing Machines (NTM). The system distinguishes between high-volume legitimate traffic and attacked traffic using learned features. Additionally, it continuously adapts to new attack patterns through regular updates, ensuring robust and up-to-date protection against evolving threats.

The hybrid nature of this model makes it particularly well-suited for the complex task of distinguishing between normal traffic, DoS attacks, and DDoS attacks in real-time network environments. Its ability to learn and remember complex patterns over time gives it an edge in detecting both sudden and gradual changes in network behavior indicative of attacks.

*B. Data Collection and Preprocessing*

*1) Dataset Description*: This study utilizes three primary datasets to evaluate network traffic and detect DoS and DDoS attacks:

*a) UNSW Normal Traffic Dataset:* This dataset includes a variety of normal network traffic patterns.

*b) Bot-IoT DoS Dataset:* This dataset contains traffic data from various DoS attacks simulated in an IoT environment.

*c) Bot-IoT DDoS Dataset:* Similar to the DoS dataset, this one includes data from DDoS attacks, also simulated in an IoT setup.

Each dataset consists of multiple features including IP addresses, port numbers, timestamps, and various protocol-specific attributes.

*2) Data Chunking and Balancing:* To ensure balanced data representation, we divided the datasets into chunks, each containing 80,000 samples. These chunks then merged and shuffled to create a mixed dataset, balancing the number of normal, DoS, and DDoS samples.

*3) Data Normalization:* All features normalized using Min-Max scaling to ensure they fall within a 0-1 range. This helps in speeding up the convergence of the neural network. Mathematically, the Min-Max scaling given by,

$$X' = \frac{X - X_{min}}{X_{max} - X_{min}} \quad (1)$$

where X is the original feature value, X' is the normalized value, Xmin is the minimum value, Xmax is the maximum value of the features.

*4) Label Encoding:* Categorical labels converted into numerical format using one-hot encoding. This step is essential for training machine learning models, particularly neural networks.

*C. Data Windowing*

*1) Sliding Window Technique:* A sliding window technique with a window size of 10-time steps applied to create sequences of data points. This process transforms the data into a 3D input shape suitable for recurrent neural networks: (samples, time steps, features).

*2) Label Adjustment:* The labels aligned with the windowed data to ensure that each sequence has a corresponding label representing the attack type or normal traffic

*D. Model Architecture*

*1) GRU Layers:* The model architecture starts with two GRU (Gated Recurrent Unit) layers. The first GRU layer consists of 64 units and is configured to return sequences to pass data to the next GRU layer. The second GRU layer has 32 units.

The GRU cell can be defined mathematically as follows:
$$z_t = \sigma(W_z \cdot [h_{t-1}, x_t]) \quad (2)$$
$$r_t = \sigma(W_z \cdot [h_{t-1}, x_t])$$
$$\tilde{h}_t = tanh(W_h \cdot [r_t \circ h_{t-1}, x_t])$$
$$h_t = (1 - z_t) \circ h_{t-1} + z_t \circ \tilde{h}_t$$

Where $z_t$ is the update gate, $r_t$ is the reset gate, $\tilde{h}_t$ is the candidate hidden state, and $h_t$ is the new hidden state.

*2) Neural Turing Machine (NTM) Layer:* A Neural Turing Machine layer integrated with the following configurations:
- Memory size and vector dimensions are defined to manage the complexity of the tasks.
- The GRU controller within the NTM processes the sequences, utilizing read and write head mechanisms to interact with the memory.

Mathematically, the NTM is composed of a controller GRU($x_t$) and a differentiable memory bank M. The read and write operations are performed as follows:
$$w_t^r = softmax(K(M, k_t^r)) \quad (3)$$
$$r_t = \sigma(W_z \cdot [h_{t-1}, x_t])$$
$$r_t = M^T w_t^r$$
$$w_t^w = softmax(K(M, k_t^w))$$
$$M_t = M_{t-1} + w_t^w \cdot v_t$$

Where K is a similarity measure, $k_t^r$ and $k_t^w$ are read and write keys, $v_t$ is the write vector, and $w_t^r$ and $w_t^w$ are read and write weights, respectively.

*3) Dense Layers*

Post the NTM layer, the model includes:
- A dense layer with 16 units using ReLU activation function:
$$ReLU(x) = max(0, x) \quad (4)$$
- An output layer with SoftMax activation to classify the traffic into distinct categories (normal, DoS, DDoS):
$$Softmax(X_i) = \frac{e^{x_i}}{\sum_j e^{x_j}} \quad (5)$$

### E. Model Compilation

*1) Loss Function:* The model uses Categorical Cross-Entropy as the loss function, appropriate for multi-class classification problems:
$$Loss = \sum_i y_i log(\hat{y}_i) \quad (6)$$

Where $y_i$ is the true label and $\hat{y}_i$ is the predicted probability.

*2) Optimizer:* Adam optimizer with a learning rate of 0.001 is used to train the model, providing a good balance between speed and accuracy:
$$m_t = \beta_1 m_{t-1} + (1 - \beta_1) g_t \quad (7)$$
$$v_t = \beta_2 v_{t-1} + (1 - \beta_2) g_t^2$$
$$\hat{m}_t = \frac{m_t}{1 - \beta_1^t}$$
$$\hat{v}_t = \frac{v_t}{1 - \beta_2^t}$$
$$\theta_t = \theta_{t-1} - \alpha \frac{\hat{m}_t}{\sqrt{\hat{v}_t} - \epsilon}$$

Where $g_t$ is the gradient, $m_t$ and $v_t$ are the first and second moment estimates, $\alpha$ is the learning rate, and $\epsilon$ is a small constant.

### F. Training Process

The training process begins with data reduction, where we randomly select 20% of the training data to manage computational resources and prevent overfitting. The training uses a batch size of 16 and runs for up to 20 epochs. We employ early stopping with a patience of 4 epochs to stop training if the validation performance does not improve, thus avoiding overfitting and saving computational resources. We also use a 20% validation split to monitor the model's performance on unseen data during training.

Callbacks are implemented to further enhance the training process. Early stopping halts training when there is no improvement in validation loss, while model checkpointing saves the best model according to validation performance. These measures ensure that the model does not overfit and generalizes well to new data. This systematic approach helps in training an effective and dependable model for practical use.

### G. Model Evaluation

In evaluating the model's effectiveness, essential metrics such as accuracy, F1-score, recall, and loss are used to evaluate its performance across both training and validation datasets. To measure the security phenomena in our research, we use several key parameters. The False Positive Rate (TPR) measures the model's ability to accurately identify attacks, while the False Positive Rate (FPR) indicates its tendency to misclassify normal traffic as attacks. Precision reflects the accuracy of positive predictions, and Recall demonstrates the model's ability to detect all attacks. The F1-score is a balanced measure of precision and recall. We also use the Area Under the ROC Curve (AUC) to determine the model's ability to distinguish between classes.

The performance of the model is evaluated using accuracy, recall, and F1-score metrics:
$$Accuracy = \frac{TP - TN}{TP + TN + FP + FN} \quad (8)$$
$$Precision = \frac{TP}{TP + FP}$$
$$Recall = \frac{TP}{TP + FN}$$
$$F1 - Score = \frac{2 \times Precision \cdot Recall}{Precision + Recall}$$

### H. Detection Process

The detection process is achieved through real-time analysis of network traffic using our trained GRU-NTM model. As network packets arrive, the model extracts relevant features and analyzes them in sequence. The GRU layers capture short-term patterns while the NTM component identifies long-term dependencies, allowing the model to classify each traffic instance as normal, DoS, or DDoS. High accuracy is achieved through careful data preprocessing and feature selection, the hybrid architecture combining GRU and NTM to capture both short-term and long-term patterns, extensive training on diverse, high-quality datasets, and regular model validation and fine-tuning. To ensure reliability, we employ cross-validation during training, test on multiple diverse datasets, continuously monitor and retrain with new data, implement confidence thresholds for classifications, and conduct regular performance audits in real-world environments.

## IV. RESULT ANALYSIS

### A. Confusion Matrix

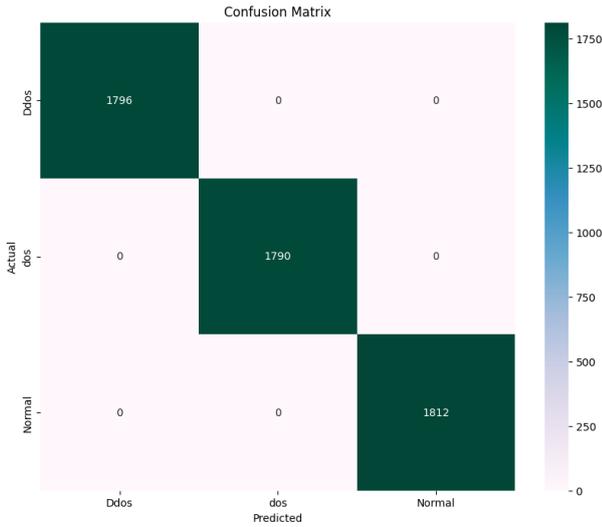

Fig. 2. Confusion Matrix

The analysis of the confusion matrix (Figure 2) serves as a critical evaluation of the model's performance in classifying DDoS, DoS, and Normal network traffic. Notably, the matrix demonstrates impeccable classification accuracy with zero misclassifications observed across all categories. Specifically, the model correctly identifies 1796 instances of DDoS attacks, 1790 instances of DoS attacks, and 1812 instances of Normal traffic.

### B. Receiver Operating Characteristic (ROC) Curve

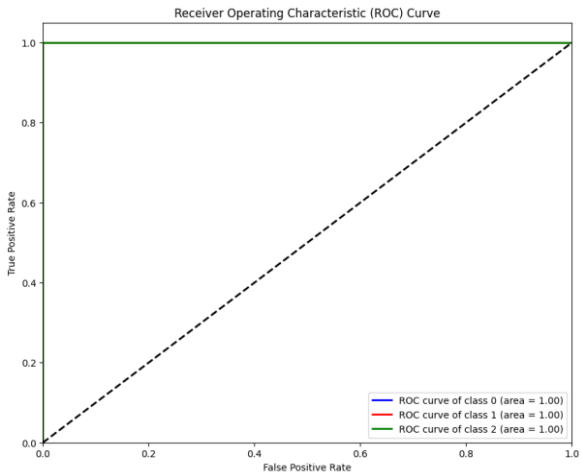

Fig. 3. ROC Curve

The Receiver Operating Characteristic (ROC) curve illustrates the optimal performance of the model across all three traffic classes: DDoS, DoS, and Normal traffic (Figure 3). The curves representing each class (blue, orange, and green) converge at the top-left corner of the plot, signifying near-perfect classification capability. The Area Under the Curve (AUC) for each class is 1.00, which indicates that the model exhibits perfect discrimination ability. This means that the model can effectively distinguish between different types of network traffic at any threshold setting, demonstrating its robustness and reliability in accurately identifying and classifying network intrusion.

### C. Training-Validation Accuracy

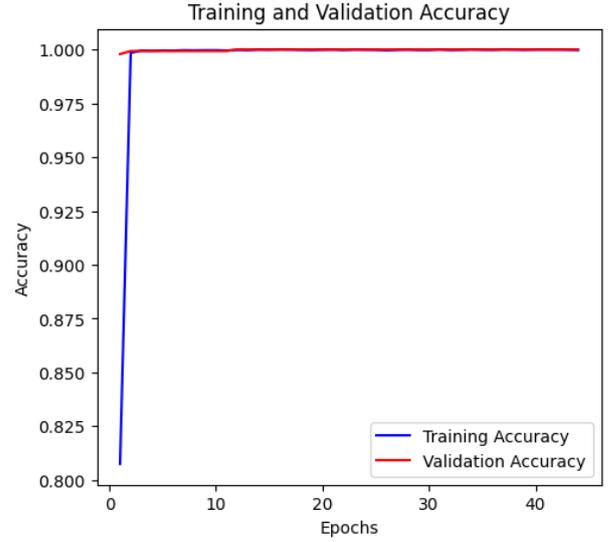

Fig. 4. Training and Validation Accuracy

- Accuracy: Above 99%
- Improved from 0.9606 (Epoch 1) to 0.9994 (Epoch 6)
- Validation accuracy reached 0.9996 by the final epoch.
- The model achieved high performance within 6 epochs.

As seen in Fig. 4, Our model's 99% accuracy is in line with top-performing models such as Coscia et al.'s [1] Anomaly2Sign (99.7% - 99.9%), Alfatemi et al.'s [2] Deep Residual Neural Networks (99.98%), and Al-zubidi et al.'s CNN-LSTM-XGBoost (98.3% - 99.3%). It slightly outperforms some models like Zeeshan et al.'s PB-DID (96.3%) and Ali et al.'s SVM (95.5%).

### D. Training-Validation Loss

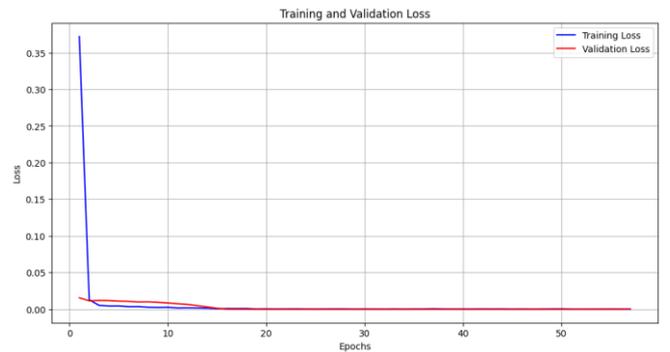

Fig. 5. Training and Validation Loss

- Training loss decreased from 0.0982 to 0.0028.
- Validation loss reduced to 0.0024 in the final epoch.

In Figure 5, the loss plot reinforces the accuracy findings. Both training and validation loss decrease rapidly in the initial epochs, converging to exceptionally low values near 0 and remaining stable. The close alignment between training and

validation loss further indicates good generalization and the absence of overfitting

*E. Implementation*

A Next-Generation Intrusion Prevention System (NGIPS) [28] is a security device that can detect vulnerabilities and provide active protection for a network. By integrating our proposed GRU-NTM model, an NGIPS can enhance its capability to detect and prevent DoS and DDoS attacks.

In real-time traffic analysis, the NGIPS would continuously monitor network traffic and feed the data through the GRU-NTM model. The model's ability to process sequential data makes it well-suited for analyzing ongoing traffic patterns and detecting anomalies. For vulnerability detection, the GRU layers in the model can identify short-term anomalies in traffic that might indicate an ongoing attack or vulnerability exploitation. Meanwhile, the NTM component can recognize subtle, long-term patterns suggesting persistent threats or slow-developing attacks.

Upon detecting potential DoS or DDoS attacks, the NGIPS can take several active protection measures. These include automatically blocking or redirecting traffic from suspected malicious sources, imposing bandwidth restrictions on suspicious traffic, monitoring and limiting connections from single sources, and enforcing protocol compliance by dropping non-compliant packets.

The adaptive nature of the GRU-NTM model allows the NGIPS to continuously refine its protection strategies. As new attack patterns emerge, the model updates its understanding, enabling the NGIPS to adjust its defenses accordingly. The NTM's memory component also maintains a record of network behavior over time. This feature is invaluable for post-incident forensic analysis and identifying the root causes of detected vulnerabilities.

The NGIPS can also share its insights with other security systems, such as firewalls and SIEM (Security Information and Event Management) systems, creating a more integrated security ecosystem.

The adaptive nature of the GRU-NTM model allows the NGIPS to continuously refine its protection strategies. As new attack patterns emerge, the model updates its understanding, enabling the NGIPS to adjust its defenses accordingly.

*F. Discussion*

Our proposed GRU-NTM hybrid model achieves an accuracy of 99% in distinguishing between normal, DoS, and DDoS traffic. This performance is similar to many of the advanced methods mentioned in the literature review:

*1) Accuracy:* Our model's 99% accuracy is in line with top-performing models such as Coscia et al.'s Anomaly2Sign (99.7% - 99.9%) [1], Alfatemi et al.'s Deep Residual Neural Networks (99.98%) [2], and Al-zubidi et al.'s CNN-LSTM-XGBoost (98.3% - 99.3%) [4]. It slightly outperforms some models like Zeeshan et al.'s PB-DID (96.3%) [6] and Ali et al.'s SVM (95.5%) [8].

*2) Model Complexity:* Our GRU-NTM hybrid model offers a unique approach compared to other hybrid models like CNN-LSTM [3] or CNN-LSTM-XGBoost [4]. The incorporation of the Neural Turing Machine allows for potentially better handling of long-term dependencies in network traffic patterns.

*3) Feature Selection:* Unlike some approaches that focus heavily on feature reduction [6][12], our model works with a comprehensive set of features. This could potentially provide more robust detection across various attack types.

*4) Real-time Performance:* While our study does not explicitly test real-time performance, the high accuracy and the nature of RNNs suggest potential for real-time application, similar to models like lightweight IDS [5] or real-time detection system [12].

*5) Novelty:* Our GRU-NTM approach represents a novel architecture in this field. While it doesn't reach the highest reported accuracy (e.g., Al-Eryani et al.'s 99.99% with Gradient Boosting [15]), it offers a new direction for research in DDoS detection.

*6) Adaptability:* The memory component of our NTM potentially allows for better adaptation to evolving attack patterns, addressing a limitation noted in several studies [10][15]

V. CONCLUSION

*A. Conclusion*

The exceptional performance of the GRU-NTM hybrid model in network intrusion detection has significant implications for cybersecurity and opens avenues for future research. The model's rapid convergence and high accuracy (>99% across all metrics) suggest its potential for real-time threat detection, potentially reducing response latency in network security systems. The near-perfect classification accuracy implies a substantial reduction in false positives, which could optimize resource allocation in network management. Furthermore, the model's ability to discriminate between DoS and DDoS attacks with high precision indicates potential for more nuanced, attack-specific response strategies. However, to fully realize these benefits, several key areas warrant further investigation.

*B. Future Enhancements*

To further improve the model's performance, several strategies can be implemented. These include fine-tuning hyperparameters techniques such as grid search or Bayesian optimization, implementing ensemble methods that combine our GRU-NTM with other models, regularly updating the training data to include new attack patterns, and optimizing the model architecture by potentially adding or removing layers. To enhance precision, we propose addressing class imbalance in the training data, implementing cost-sensitive learning, fine-tuning the decision threshold for each class, incorporating domain-specific features that better distinguish attack traffic, and regularly updating the model with the latest attack signatures. These enhancements aim to not only improve the model's accuracy and precision but also its adaptability to new and evolving attack patterns.

The development of interpretability methods for the GRU-NTM architecture could provide constructive insights into the model's decision-making method, potentially uncovering novel attack signatures or network vulnerabilities. Investigation into transfer learning capabilities could enable rapid adaptation to emerging threat landscapes. Furthermore, subjecting the model to adversarial testing regimes would be instrumental in identifying and mitigating potential vulnerabilities, thereby enhancing its robustness. Finally, research into privacy-preserving implementations, such as federated learning, could address data sensitivity concerns in multi-stakeholder network environments. These enhancements aim to not only validate the model's effectiveness across various scenarios but also to address potential limitations, ultimately resulting in the development of more adaptive and resilient network security paradigms.